\documentclass[twoside,twocolumn,english,aps,pra,superscriptaddress]{revtex4-1}
\usepackage[T1]{fontenc}
\usepackage[latin9]{inputenc}
\usepackage{textcomp}
\usepackage{amsmath}
\usepackage{amssymb}
\usepackage{graphicx}
\usepackage{xcolor}
\usepackage{soul} 
\usepackage{ulem} 

\makeatletter
\@ifundefined{textcolor}{}
{%
 \definecolor{BLACK}{gray}{0}
 \definecolor{WHITE}{gray}{1}
 \definecolor{RED}{rgb}{1,0,0}
 \definecolor{GREEN}{rgb}{0,1,0}
 \definecolor{BLUE}{rgb}{0,0,1}
 \definecolor{CYAN}{cmyk}{1,0,0,0}
 \definecolor{MAGENTA}{cmyk}{0,1,0,0}
 \definecolor{YELLOW}{cmyk}{0,0,1,0}
}

\makeatother

\usepackage{babel}
\begin{document}


\title{Long-time efficacy of the surface code in the presence of a
  superohmic environment}

\author{D. A. L\'opez-Delgado}

\affiliation{Departamento de F\'{\i}sica da Mat\'eria Condensada,
  Instituto de F\'{\i}sica Gleb Wataghin, Universidade Estadual de
  Campinas, Campinas, S\~ao Paulo 13083-970, Brazil}

\author{E. Novais}

\affiliation{Centro de Ci\^encias Naturais e Humanas, Universidade
  Federal do ABC, Santo Andr\'e, S\~ao Paulo 09210-170, Brazil}

\author{E. R. Mucciolo}

\affiliation{Department of Physics, University of Central Florida,
  Orlando, Florida 32816, USA}

\author{A. O. Caldeira}

\affiliation{Departamento de F\'{\i}sica da Mat\'eria Condensada,
  Instituto de F\'{\i}sica Gleb Wataghin, Universidade Estadual de
  Campinas, Campinas, S\~ao Paulo 13083-970, Brazil}


\begin{abstract}
We study the long-time evolution of a quantum memory coupled to a bosonic
 environment on
which  quantum error correction (QEC) is performed using the surface code. 
The memory's evolution encompasses $N$ QEC cycles, each of them
yielding a non-error syndrome. This assumption makes our analysis
independent of the recovery process.
We map the expression for the time evolution of the memory onto the partition
function of an equivalent statistical mechanical spin system.
In the superohmic dissipation case the long-time evolution of the memory has
the same behavior as the time evolution for just one QEC cycle.
For this case we find analytical expressions for the critical parameters of the
order-disorder phase transition of an equivalent spin system.
These critical parameters
determine the threshold value for the system-environment coupling below which
it is possible to preserve the memory's state.
\end{abstract}

\maketitle

\section{Introduction}

It is widely accepted that large-scale quantum information processing
will demand some sort of quantum error correction (QEC) as a
fundamental part of its design
\cite{DiVincenzo.2009,Nature.506.2014,QECBook}. The standard analysis
of QEC codes deals with their efficiency against stochastic errors
\cite{Bravyi2010,PhysRevA.86.032324}. Although stochastic error models
can sometimes be justified on physical grounds, often they are used for
their simplicity rather than their accuracy in describing the effect
of realistic environments. Thus, it is important to complement these
studies to include errors caused by environments that can be
microscopically modeled
\cite{NMB07,PhysRevA.79.032318,PhysRevLett.110.010502,Pre13}. In
particular, Gaussian bosonic environments are amenable to analytical
and numerical studies and have characteristics that are beyond the
standard stochastic models, such as correlations and memory effects.

The surface code is regarded as a paradigmatic QEC code
\cite{SurfCode,DKL+02,DiVincenzo.2009,ISI:000272310000047,ISI:000296073000005,FWH12}.
It only requires local gates and has a large ``threshold'' against
stochastic error. In addition to the usual stochastic error threshold,
it has also been partially benchmarked against Gaussian error models
\cite{PhysRevLett.110.010502, PhysRevA.88.012336, PhysRevA.90.042315,
  PhysRevX.4.031058, arXiv:1606.0316}.  Unfortunately, all previous
work focused only on the quantum information fidelity after a single
QEC cycle.

In this paper, we take the discussion of the efficiency of the surface
code against Gaussian noise one step forward. We provide analytical
expressions for the logical qubit fidelity after an arbitrary number
of QEC cycles in the presence of a bosonic Gaussian environment. We
then specialize the calculation for a particular ``superohmic''
environment and demonstrate that there are two possible regimes,
depending on whether the coupling between the qubits and the
environment is below or above a critical value: (i) below the critical
value, information is preserved by QEC; (ii) information is lost
otherwise. We further demonstrate that the ``threshold'' for
a superohmic environment is identical to the ``threshold'' for a
single QEC cycle. Hence, we demonstrate that for a superohmic
environment memory effects and correlations between QEC cycles are
unimportant, thus confirming an old conjecture for a dense set of
qubits \cite{NMB10}.

We start our discussion in Sec. \ref{sec:assumptions} by describing
all the assumptions built into our calculation. In
Sec. \ref{sec:Fidelity}, we derive the analytical expression for the
fidelity after many QEC cycles and specialize the calculation for a
superohmic environment in Sec.
\ref{sec:Super-ohmic-environment-with}. Finally, we present our
conclusions and perspectives in Sec. \ref{sec:Conclusions}.

\section{The surface code in a Gaussian environment}
\label{sec:assumptions}

All threshold analyses of QEC are based on some assumptions about the
system and its environment. We start by presenting all the assumptions
built into our expressions for the logical qubit fidelity in a concise
and itemized fashion to make the text clear and accessible. Our
assumptions are in general favorable to the success of QEC. Therefore,
our results must be regarded as a upper bound to the efficiency of the
code against Gaussian noise. We follow the list of assumptions with a
brief review of the surface code in Sec. \ref{sec:Surface-code}, and
finally introduce the Gaussian bosonic environment in Sec.
\ref{sec:environment}.

The following assumptions relate to the ability to manipulate qubits:
\begin{itemize}
\item We assume that the initial logical state can be prepared
  flawlessly and it is disentangled from the environment. Since this
  is one of the DiVincenzo's criteria \cite{NC00-a}, we do not regard
  this as a fundamental limitation to the analysis.
\item In order to avoid additional assumptions about how quantum gates
  are performed \cite{Novais:2008:012314}, we focus on a quantum
  memory. In other words, after quantum information is encoded into
  the logical Hilbert space, no quantum gate is performed.
\item In a real situation, syndrome extraction would be faulty and
  time consuming, and could excite the environment
  \cite{DiVincenzo:2007:020501}. Thus, in a real situation, some
  modelling of the measurement apparatus would have to be
  considered. In order to avoid this extra layer of complexity, we opt
  for considering the syndrome extraction to the flawless and
  instantaneous.
\item We derive a general expression for the fidelity of the logical
  qubit after several QEC steps. At the end of each cycle different
  syndromes could be measured and a proper recovery operation would
  have to be performed \cite{DP10}. However, there is no guarantee
  that the recovery operation would be the correct one
  \cite{PhysRevA.90.042315}. It is only possible to define an
  intrinsic threshold for the code if we assume that all syndrome
  measurements return a nonerror (more precisely, no detectable
  errors). Even though this is a very particular evolution, it is the
  only one that does not depend on a recovery strategy.
\end{itemize}

There are several possible microscopic models that apply to real
environments. However, very few choices are amenable to an analytical
or numerical calculation. In order to gain some insight into the basic
structure of a quantum environment and its effects on the surface code
threshold, we assume that:
\begin{itemize}
\item The environment can be initially refrigerated to its lowest
  possible energy state. It is conceivable that over a long period of
  time an environment can be refrigerated to extremely low
  temperatures. This constitutes the very best scenario for quantum
  information processing, since it provides the best possible
  coherence times. However, more likely in practice, after this
  initial step, the dynamics between the qubits and the environment
  can lead to excitations in the environment. We thus assume that the
  duration of the QEC is shorter than the time needed to refrigerate
  the environment.
\item In order to derive exact analytical expression for the evolution
  operator of the qubits, we restrict the errors induced by the
  environment to bit flips.
\item There are many possible dispersion relations for a Gaussian
  environment. A particularly simple choice is to consider a linear
  dispersion relation, $\omega_{k}=v\left|{\bf k}\right|$, and a
  constant velocity of excitations, $v$. This is not a restricting
  choice, but just a convenient one. A more crucial quantity to be
  defined is the environment's spectral function
  \cite{CL83,breuer2007theory,CaldeiraBook}.
\item It is natural to assume the existence of a large cutoff
  frequency for the environmental modes, $\omega_{\Lambda}$. Although
  the ultraviolet cutoff can be a large number, physical
  characteristics of the system and the enviroment make it
  finite. Typically, form factors in the coupling between the qubits
  and the environment define an ultraviolet scale to the system. A
  simple example is a charge qubit in a double quantum dot. The
  highest frequency phonon that couples to this qubit is not of the order
  of the Debye frequency, but it is set by the inverse of the distance
  between the dots \cite{Vorojtsov:2005:205322}.
\end{itemize}
%

\subsection{Surface code}
\label{sec:Surface-code}

\begin{figure}
\begin{centering}
\includegraphics[scale=0.36]{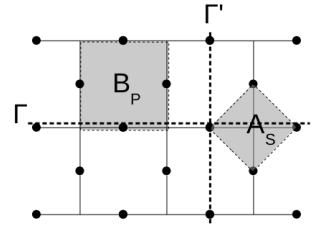}
\par\end{centering}
\caption{The surface code. Physical qubits are shown as black
  dots. The stabilizer operators are products of four Pauli operators,
  $A_{S}$ and $B_{P}$. Logical operators are correspond to string of
  Pauli operators crossing the qubit lattice from one side to the
  opposite, $\Gamma$ and $\Gamma'$.}
\label{fig:surf-code}
\end{figure}

The physical qubits in the surface code are arranged on the edges of a
square lattice, forming themselves a square lattice slanted by
45$^{\rm o}$. The QEC scheme is based on the stabilizer formalism
\cite{NC00-a} with two sets of stabilizer: plaquettes
\begin{equation}
B_{P}=\prod_{i\in P}\sigma_{i}^{z},
\end{equation}
and stars
\begin{equation}
A_{S}=\prod_{i\in S}\sigma_{i}^{x},
\end{equation}
operators, as shown in Fig. (\ref{fig:surf-code}).

When all stabilizers are enforced, the lattice of physical qubits in
Fig. \ref{fig:surf-code} encodes one logical qubit. The logical bit
flip operator, $\bar{X}$, is the product of physical bit flip
operators, $\sigma^{x}$, along a path from the upper to the lower
sides of the lattice, e.g., through path $\Gamma$ in
Fig. (\ref{fig:surf-code}). Similarly, a logical phase flip operator,
$\bar{Z}$, is the product of physical phase flips
operators,$\sigma^{z}$, joining the vertical boundaries of the
lattice, e.g., path $\Gamma'$, in Fig. \ref{fig:surf-code}.

Additional logical qubits can be encoded in the same set of physical
qubits by relaxing the stabilizers constrains
\cite{PhysRevA.86.032324}. However, the case of a single logical qubit
is expected to have the largest possible threshold. Hence, we focus on
this situation and define the logical states as
\begin{equation}
| \bar{\uparrow} \rangle = \frac{1}{\sqrt{N_{\lozenge}}} G | F \rangle
\end{equation}
and
\begin{equation}
| \bar{\downarrow} \rangle = \bar{X}| \bar{\uparrow} \rangle,
\end{equation}
where $|F\rangle$ is the ferromagnetic $z$ state of the physical
qubits, $N_{\lozenge}$ is a normalization constant, $G =
\prod_{\lozenge} \left( 1+A_{\lozenge} \right)$, and
$\prod_{\lozenge}$ is the product over all star stabilizers.

\subsection{Environment}
\label{sec:environment}

The traditional ``system plus environment'' approach to open quantum
systems \cite{RevModPhys1987,CaldeiraBook} is the most natural method
to systematically study memory effects and spatial correlations. In
this approach, the environment is described
by a large set of harmonic oscillators,
\begin{equation}
H_{0}=\sum_{{\bf k}}\omega_{{\bf k}}a_{{\bf k}}^{\dagger}a_{{\bf k}},
\end{equation}
where $a_{\mathbf{k}}$ and $a_{\mathbf{k}}^{\dagger}$ are bosonic
annihilation and creation operators that have the usual canonical
commutation relations, i.e., $\left[ a_{\mathbf{k}},
  a_{\mathbf{k'}}^{\dagger} \right] = \delta_{\mathbf{k},\mathbf{k'}}$,
where we set $\hbar=1$. The environmental modes have momentum $k_{i} =
\frac{2\pi}{L}$ $n_{i}$ with $n_{i}$ an integer and $L$ a macroscopic
characteristic length. Finally, following our assumptions we consider
the dispersion relation, $\omega_{{\bf k}}=v\left|{\bf k}\right|$.

The full quantum evolution is given by the Hamiltonian
\begin{equation}
H = H_{0} + V,
\end{equation}
where $V$ is the interaction between the qubits and the environment. A
pure bit-flip error model has the interaction Hamiltonian
\cite{Weiss:1999:1,breuer2007theory},
\begin{equation}
V = \lambda \sum_{{\bf r}} f\left(\mathbf{r}\right) \sigma_{{\bf
    r}}^{x},
\label{eq:V}
\end{equation}
where $f\left(\mathbf{r}\right)$ is the bosonic operator
\begin{equation}
f\left(\mathbf{r}\right) =
\frac{\left(v/\omega_{0}\right)^{D/2+s}}{L^{D/2}}
\sum_{\mathbf{k}\neq0} \left| \mathbf{k} \right|^{s}
\left(e^{+i\mathbf{k}\cdot\mathbf{r}}a_{\mathbf{k}}^{\dagger} +
\mbox{H. c.}\right),
\label{eq:f(r,t)}
\end{equation}
$\omega_{0}$ is a characteristic microscopic frequency scale, and $D$
the number of spatial dimensions of the environment. The power $s$
defines the low-frequency behavior of the enviroment's spectral
density, as we show below.

It is straightforward to derive the time evolution operator in the
interaction picture after a time $\Delta$,
\begin{equation}
  \hat{U}\left(\Delta,0\right) = T_{t} e^{-i\lambda\int_{0}^{\Delta}dt
    \sum_{\mathbf{r}} \hat{f}
    \left(\mathbf{r},t\right)\sigma_{\mathbf{r}}^{x}},
  \label{eq:U(delta)}
\end{equation}
where 
\begin{equation}
\hat{f} \left(\mathbf{r},t\right) = \frac{1}{L^{D/2}}
\sum_{\mathbf{k}\neq0} \left|\mathbf{k}\right|^{s}
\left(e^{+i\mathbf{k} \cdot \mathbf{r}+i\omega_{k}t}
a_{\mathbf{k}}^{\dagger} + e^{-i\mathbf{k} \cdot
  \mathbf{r}-i\omega_{k}t} a_{\mathbf{k}} \right),
\label{eq:f(r,t)-1}
\end{equation}
and $T_{t}$ is the time ordering operator.

Although the dispersion relation of the environment is an important
quantity, what traditionally defines the type of environment is its
spectral density
\begin{equation}
J(\omega) \equiv \frac{\pi}{2}\sum_{k}\frac{C_{k}^{2}}{m_{k}\omega_{k}}
\delta(\omega-\omega_{k}).
\label{eq:J(omega)}
\end{equation}
Here $C_{k}$ comes from the interaction term, Eq. (\ref{eq:V}), when it is
written in the  coordinate-coordinate coupling scheme,
$V=q_{0}\sigma_{x}\sum_{k}C_{k}q_{k}$. 
In the latter,  the length scale $q_0$ is introduced in such a way we recover
the appropriate dimension of the interaction term. Usually, it comes from the
original Hamiltonian of the system we are representing here as our physical 
qubits. For example, $\pm q_0$ could be the positions of two degenerate minima 
of a localized tunneling center \cite{breuer2007theory}. 

In order to identify $C_k$ for our model, we use Eqs. (\ref{eq:V}) and
(\ref{eq:f(r,t)}), and $q_k = \sqrt{1/2m_{k}\omega_{k}}(a_{k}+a_{k}^{\dagger})$.
In this way we get:
\begin{equation}
C_{k}^{2} = 2\lambda^{2}\frac{(v/\omega_{0})^{D+2s}}{q_{0}^{2}L^{D}}
    m_{k}vk^{2s+1}.
\end{equation}

Now we can calculate the spectral density. We do this by inserting $C_k^2$ 
into Eq. (\ref{eq:J(omega)}), and taking the continuum limit
(i.e. $\frac{(2\pi)^{D}}{L^{D}}\sum_{k} \rightarrow \int d^{D}k$). Finally
we assume a two-dimensional environment, i.e. $D=2$, and
perform the corresponding integration, obtaining:
\begin{equation}
J(\omega) = \frac{2\lambda^{2}}{q_{0}^{2}\omega_{0}^{2+2s}}\omega^{2s+1}.
\label{eq:J(omega)-1}
\end{equation}

Thus, following the standard definition, a two-dimensional environment
with $s<0$ is known as subohmic, while for $s=0$ and $s>0$ they are
known as ohmic and superohmic, respectively.

We can use some of the properties of Gaussian environments to simplify
Eq. (\ref{eq:f(r,t)-1}) \cite{arXiv:1606.0316}. We first use the
Magnus expansion \cite{Blanes2009151} for the evolution operator to
deal with the time ordering operator and then normal order the
exponentials, finally arriving at
\begin{equation}
  \hat{U} \left(\Delta,0\right) = \left[\prod_{{\bf k}\neq0}
    e^{-\hat{{\cal G}}\left({\bf k},\Delta,0\right)}\,
    e^{-i\hat{\alpha}\left({\bf k},\Delta,0\right)a_{{\bf
          k}}^{\dagger}} e^{-i\hat{\alpha}^{*}\left({\bf
        k},\Delta,0\right)a_{{\bf k}}} \right],
  \label{eq:evol-1}
\end{equation}
where

\begin{equation}
\hat{\alpha}\left({\bf k},\Delta,0\right) = 
\frac{\lambda}{L^{D/2}}\int_{\Delta\left(n-1\right)}^{\Delta n}dt 
\sum_{\mathbf{r}}|\mathbf{k}|^{s}\hat{\sigma}_{\mathbf{r},n}^{x}
e^{+i\mathbf{k}\cdot\mathbf{r}+i\omega_{k}t} ,
\label{eq:alpha}
\end{equation}
and
\begin{eqnarray}
\hat{{\cal G}}\left({\bf k}\right) & = & - \frac{\lambda^{2}}{L^{D}}
\left(v/\omega_{0}\right)^{D+2s} \int_{0}^{\Delta}
dt_{1}\int_{0}^{\Delta}dt_{2} \theta \left(t_{1}-t_{2}\right)
\nonumber \\ & & \times\ \sum_{\mathbf{r},\mathbf{r'}}
\left|\mathbf{k}\right|^{2s} e^{-i\mathbf{k} \cdot
  \left(\mathbf{r}-\mathbf{r'}\right) -
  i\omega_{k}\left(t_{1}-t_{2}\right)} \hat{\sigma}_{\mathbf{r}}^x
\hat{\sigma}_{\mathbf{r'}}^x .
\label{eq:calG}
\end{eqnarray}
A detailed derivation of these steps can be found in
Ref. \cite{arXiv:1606.0316}.

For later convenience, we rewrite Eq. (\ref{eq:evol-1}) in the qubit
$x$ basis $\{|\pm\rangle\}$, where $\hat{\sigma}^{x}|\pm\rangle =
\pm|\pm\rangle$. By defining $\left|\sigma\right\rangle$ as a
configuration of qubits with eigenvalue $\sigma_{{\bf r}}^x = \pm 1$
for the qubit at position ${\bf r}$, we recast Eq. (\ref{eq:evol-1})
as
\begin{equation}
  \hat{U} \left(\Delta,0\right) = \sum_{\bar{\sigma}}
  u\left(\bar{\sigma}\right) \left| \bar{\sigma} \right\rangle
  \left\langle \bar{\sigma} \right|,
  \label{eq:u-bosonic+spins}
\end{equation}
where $\bar{\sigma}=\{\sigma_{\mathbf{r}}^x\}$ denotes the full set of
spin variables, the pure bosonic operator is defined as
\begin{equation}
  u \left( \bar{\sigma} \right) = \prod_{{\bf k}\neq0} e^{-{\cal
      G}\left({\bf k}\right)}\, e^{-i\alpha\left({\bf k}\right)\,
    a_{{\bf k}}^{\dagger}}\, e^{-i\alpha^{*}\left({\bf k}\right)\,
    a_{{\bf k}}},
  \label{eq:u-bosonic}
\end{equation}
and the auxiliary functions are given by
\begin{eqnarray}
{\cal G}\left({\bf k}\right) & = & - \frac{\lambda^{2}}{L^{D}}
\left(v/\omega_{0}\right)^{D+2s} \int_{0}^{\Delta}
dt_{1}\int_{0}^{\Delta}dt_{2} \theta \left(t_{1}-t_{2}\right)
\nonumber \\ & & \times\ \sum_{\mathbf{r},\mathbf{r'}}
\left|\mathbf{k}\right|^{2s} e^{-i\mathbf{k} \cdot
  \left(\mathbf{r}-\mathbf{r'}\right) -
  i\omega_{k}\left(t_{1}-t_{2}\right)} \sigma_{\mathbf{r}}^x
\sigma_{\mathbf{r'}}^x,
\label{eq:calG-bosonic}
\end{eqnarray}
\begin{equation}
\alpha\left({\bf k},\Delta,0\right) = 
\frac{\lambda}{L^{D/2}}\int_{\Delta\left(n-1\right)}^{\Delta n}dt 
\sum_{\mathbf{r}}|\mathbf{k}|^{s}\sigma_{\mathbf{r},n}^{x}
e^{+i\mathbf{k}\cdot\mathbf{r}+i\omega_{k}t} .
\label{eq:alpha-bosonic}
\end{equation}
%

\section{Logical Qubit Fidelity}
\label{sec:Fidelity}

We initially set the system in the logical state
$|\bar{\uparrow}\rangle$ and and assume that the environment is in its
ground state $|0\rangle$,
\begin{equation}
  |\psi_{0}\rangle = |\bar{\uparrow} \rangle \otimes| 0\rangle.
  \label{eq:init-state}
\end{equation}
We consider that the syndromes are extracted at equal time intervals
of duration $\Delta$ and that the extraction is performed
instantaneously. Hence, the full evolution operator is composed by a
sequence of unitary evolutions and projections. In the simplest case
of a nonerror syndrome, where there is no recovery operation to be
performed, the quantum state after $N$ cycles is given by
\begin{eqnarray}
|\psi\rangle & = & \mathcal{P}_{0}\, U \left(N\Delta,\left(N-1\right)
\Delta\right) \ldots \nonumber \\ & & \ldots \mathcal{P}_{0}\, U
\left(2\Delta,\Delta\right) \mathcal{P}_{0}\, U\left(\Delta,0\right)|
\psi_{0} \rangle,
\label{eq:evolstate}
\end{eqnarray}
where
\begin{equation}
  \mathcal{P}_{0} = |\bar{\uparrow} \rangle\langle \bar{\uparrow}| +
  \bar{X}| \bar{\uparrow }\rangle\langle \bar{\uparrow}| \bar{X}.
  \label{eq:projection-oper}
\end{equation}
We now introduce a subscript to the spin variable sets to designate
the QEC step where the spin states evolve and use
Eq. (\ref{eq:u-bosonic+spins}) to rewrite
\begin{eqnarray}
|\psi\rangle & = & \sum_{\bar{\sigma}_{1} \ldots \bar{\sigma}_{N}}
\left[ u\left(\bar{\sigma}_{N}\right) \ldots u \left(\bar{\sigma}_{2}
  \right)\, u \left(\bar{\sigma}_{1}\right) \left| 0 \right\rangle
  \right] {\cal P}_{0} \left|\bar{\sigma}_{N} \right\rangle \nonumber
\\ & & \left\langle \bar{\sigma}_{N} \right|\mathcal{P}_{0}\left|
\bar{\sigma}_{N-1} \right\rangle \ldots \left\langle \bar{\sigma}_{2}
\right| \mathcal{P}_{0} \left| \bar{\sigma}_{1} \right\rangle
\left\langle \bar{\sigma}_{1}| \bar{\uparrow} \right\rangle,
\label{eq:evolstate-1}
\end{eqnarray}
where $\sigma_{{\bf r}}^{x}\left| \bar{\sigma}_{n} \right\rangle =
\sigma_{{\bf r},n} \left| \bar{\sigma}_{n} \right\rangle$ and
$\sigma_{{\bf r},n} = \pm 1$. Since we have already integrated over
time in Eqs. (\ref{eq:calG-bosonic}) and (\ref{eq:alpha-bosonic}),
these labels work now as new time variables.

The projectors can also be easily expressed in the $x$ basis of the
qubits,
\begin{equation}
\mathcal{P}_{0} = \frac{1}{2^{M}N_{\lozenge}}
\sum_{\bar{\sigma},\bar{\sigma}'} \sum_{{\cal J}=\left\{
  \bar{I},\bar{X}\right\} } G{\cal J}| \bar{\sigma} \rangle \langle
\bar{\sigma}'| {\cal J} G
\end{equation}
where $M$ the total number of qubits and $\left\{
\bar{\sigma},\bar{\sigma}' \right\} $ are two independent sum of all
qubit configurations in the $x$ basis. After relabeling the indexes
and using the property $G^{2}=N_{\lozenge}G$, we obtain
\begin{eqnarray}
  |\psi\rangle & = & \frac{1}{\sqrt{{\cal N}}} \sum_{\left\{
    \bar{\sigma}_{i}, {\cal J}_{i}\right\} } \left[ \prod_{n=0}^{N-1}
    \nonumber u \left( \bar{\sigma}_{n}\right) \left|0 \right\rangle
    \right] \\ & & \times\, {\cal J}_{N-1}\, G \left|
  \bar{\sigma}_{N-1} \right\rangle \prod_{n=0}^{N-2} \left\langle
  \bar{\sigma}_{n} \right|{\cal J}_{n}\, G \left| \bar{\sigma}_{n}
  \right\rangle
\label{eq:ketpsi}
\end{eqnarray}
with ${\cal N}$ being a normalization constant and ${\cal
  J}_{i}=\left\{ \bar{I},\bar{X}\right\} $.

When we use Eq. (\ref{eq:ketpsi}) to evaluate an expectation value, we
have terms with the same ``time'' label coming from the ket and the
bra. Hence, it is convenient to differentiate the origin of each term
by renaming the variables in the bra $\left\langle \psi\right|$ as
\begin{align*}
\sigma & \to\tau,\\
\alpha & \to\beta,\\
{\cal J} & \to{\cal K}.
\end{align*}
It is now straightforward to write the logical state fidelity after
$N$ QEC cycles,
\begin{equation}
\mathcal{F} = \frac{ \left\langle \psi| \bar{\uparrow} \right\rangle
  \left\langle \bar{\uparrow}| \psi \right\rangle} {\langle\psi |
  \psi\rangle},
\label{eq:fidelity}
\end{equation}
where
\begin{eqnarray}
\left\langle \psi| \bar{\uparrow} \right\rangle \left\langle
\bar{\uparrow}| \psi \right\rangle & = & \frac{1}{{\cal N}}
\sum_{\left\{ \bar{\sigma}_{i}, \bar{\tau}_{i}\right\} } \sum_{\left\{
  {\cal J}_{i}, {\cal K}_{i} \right\} }^{\prime} \nonumber \\ & &
\times\, \left\langle 0 \right| \left[ \prod_{k=N-1}^{0}u^{\dagger}
  \left( \bar{\tau}_{k} \right) \right] \left[ \prod_{j=0}^{N-1} u
  \left( \bar{\sigma}_{j} \right) \right] \left|0\right\rangle
\nonumber \\ & & \times\, \prod_{l=0}^{N-1} \left\langle
\bar{\tau_{l}} \right| {\cal J}_{l}\, G\left| \bar{\tau}_{l}
\right\rangle \left\langle \bar{\sigma}_{l} \right|{\cal J}_{l}\,
G\left| \bar{\sigma}_{l} \right\rangle,
\end{eqnarray}
where $\sum^{\prime}$ denotes a summation with the restrictions
$\mathcal{J}_{0} = \mathcal{K}_{0} = \mathcal{J}_{N-1} =
\mathcal{K}_{N-1}=\bar{I}$. Similarly,

\begin{eqnarray}
\langle\psi|\psi\rangle & = & \frac{1}{{\cal N}} \sum_{\left\{
  \bar{\sigma}_{i}, \bar{\tau}_{i}\right\} } \sum_{\left\{ {\cal
    J}_{i},{\cal K}_{i}\right\} } \nonumber \\ & & \times\,
\left\langle 0\right| \left[ \prod_{k=N-1}^{0}u^{\dagger} \left(
  \bar{\tau}_{k} \right) \right] \left[ \prod_{j=0}^{N-1}u \left(
  \bar{\sigma}_{j} \right) \right] \left| 0 \right\rangle \nonumber
\\ & & \times\, \prod_{l=0}^{N-1} \left\langle \bar{\tau}_{l} \right|
   {\cal J}_{l}\, G\left| \bar{\tau}_{l} \right\rangle \left\langle
   \bar{\sigma}_{l} \right| {\cal J}_{l}\, G \left| \bar{\sigma}_{l}
   \right\rangle,
\end{eqnarray}
but with no restrictions on $\mathcal{J}_{n}$ or $\mathcal{K}_{n}$. 
Notice $\langle\psi|\psi\rangle \neq 0$ in general,
since the time evolution of the system is non-unitary.

To further simplify these expressions, we normal order the expectation
value of the bosonic operators. This is a tedious task, but easily
performed using the Baker-Campbell-Hausdorff formula. For instance,
\begin{eqnarray}
e^{-i\alpha^{*}({\bf k},n)\, a_{{\bf k}}}\, e^{-i\alpha({\bf k},m)\,
  a_{{\bf k}}^{\dagger}} & = & e^{-i\alpha({\bf k},m)\,a_{{\bf
      k}}^{\dagger}}\, e^{-i\alpha^{*}({\bf k},n)\,a_{{\bf k}}}
\nonumber \\ & & \times\, e^{-\alpha^{*}({\bf k},n)\,\alpha({\bf
    k},m)}.
\label{eq:baker-hausdorff}
\end{eqnarray}
After performing several commutations to normal order the bosonic
operators, we obtain
\begin{widetext}
\begin{eqnarray}
\left\langle 0 \right| \left[ \prod_{k=N-1}^{0} u^{\dagger} \left(
  \bar{\tau}_{k} \right) \right] \left[\prod_{j=0}^{N-1}u \left(
  \bar{\sigma}_{j} \right) \right] \left| 0\right \rangle & = &
\prod_{{\bf k}\neq0} \exp \left\{ -\sum_{n=0}^{N-1} \left[ {\cal
    G}({\bf k},n) + {\cal G}^{*}({\bf k},n) - \beta^{*}({\bf k},n)\,
  \alpha({\bf k},n) \right] \right\} \nonumber \\ & & \times\, \exp
\left\{ -\ \sum_{n=1}^{N-1}\sum_{m=0}^{n-1} \left[ \alpha^{*}({\bf
    k},n)\, \alpha({\bf k},m) + \beta({\bf k},n)\, \beta^{*}({\bf
    k},m) \right. \right. \nonumber \\ & &
  \left. \left. -\ \beta^{*}({\bf k},n)\, \alpha({\bf k},m) -
  \alpha({\bf k},n)\, \beta^{*}({\bf k},m) \right] \right\}.
\label{eq:product}
\end{eqnarray}
It is natural to rewrite this equation in the short and suggestive
form of an exponential,
\begin{eqnarray}
\left\langle 0\right| \left[ \prod_{k=N-1}^{0} u^{\dagger} \left(
  \bar{\tau}_{k} \right) \right] \left[ \prod_{j=0}^{N-1} u \left(
  \bar{\sigma}_{j} \right) \right] \left| 0 \right\rangle & = &
e^{-{\cal H}},
\end{eqnarray}
where
\begin{eqnarray}
{\cal H} & = & \sum_{{\bf r},{\bf s}} \Big\{ \sum_{n=0}^{N-1}
F_{1}({\bf r}-{\bf s},0)\, (\tau_{{\bf r},n} - \sigma_{{\bf r},n})
(\tau_{{\bf s},n} - \sigma_{{\bf s},n}) \nonumber \\ & & +\ i
\Phi_{1}({\bf r}-{\bf s})\, (\tau_{{\bf s},n} - \sigma_{{\bf s},n})
(\tau_{{\bf r},n}+\sigma_{{\bf r},n}) + i\Phi_{2}({\bf r}-{\bf
  s},0)(\tau_{{\bf s},n} \sigma_{{\bf r},n}-\tau_{{\bf r},n}
\sigma_{{\bf s},n}) \nonumber \\ & & -\ \sum_{n=1}^{N-1}
\sum_{m=0}^{n-1} \left[ F_{1}({\bf r}-{\bf s},n-m) - \Phi_{3}({\bf
    r}-{\bf s},n-m) \right] (\tau_{{\bf r},n} - \sigma_{{\bf r},n})
(\tau_{{\bf s},m} - \sigma_{{\bf s},m}) \nonumber \\ & & +\ i \left[
  F_{2}({\bf r}-{\bf s},n-m) + \Phi_{3}({\bf r}-{\bf s},n-m) \right]
(\tau_{{\bf r},n} - \sigma_{{\bf r},n}) (\tau_{{\bf s},m} +
\sigma_{{\bf s},m}) \Big\},
\label{eq:H}
\end{eqnarray}
and
\begin{subequations}
\label{eq:correlators}
\begin{eqnarray}
F_{1}({\bf r},n) & = &
\frac{\lambda^{2}(v/\omega_{0})^{D+2s}}{L^{D}}\sum_{{\bf k}\neq0}
\left|{\bf k}\right|^{2s}
\left[\frac{1-\cos(\omega_{k}\Delta)}{\omega_{k}^{2}}\right] \cos({\bf
  k}\cdot{\bf r}) \cos(n\omega_{k}\Delta),
\label{eq:correlator-1}
\end{eqnarray}
\begin{eqnarray}
F_{2}({\bf r},n) & = &
\frac{\lambda^{2}(v/\omega_{0})^{D+2s}}{L^{D}}\sum_{{\bf
    k}\neq0}\left|{\bf k}\right|^{2s}
\left[\frac{1-\cos(\omega_{k}\Delta)} {\omega_{k}^{2}}\right]\cos({\bf
  k} \cdot{\bf r}) \sin(n\omega_{k}\Delta), \label{eq:correlator-2}
\\ \Phi_{1}({\bf r},n) & = &
\frac{\lambda^{2}(v/\omega_{0})^{D+2s}}{L^{D}}\sum_{{\bf
    k}\neq0}\left|{\bf k}\right|^{2s} \left[\frac{\omega_{k}\Delta -
    \sin(\omega_{k}\Delta)} {\omega_{k}^{2}}\right] \cos({\bf
  k}\cdot{\bf r}),
\label{eq:correlator-3}
\end{eqnarray}
\begin{eqnarray}
\Phi_{2}({\bf r},n) & = &
\frac{\lambda^{2}(v/\omega_{0})^{D+2s}}{L^{D}} \sum_{{\bf
    k}\neq0}\left|{\bf k}\right|^{2s}
\left[\frac{1-\cos(\omega_{k}\Delta)}{\omega_{k}^{2}}\right] \sin({\bf
  k}\cdot{\bf r}) \cos(n\omega_{k}\Delta),
\label{eq:correlator-4}
\end{eqnarray}
\begin{eqnarray}
\Phi_{3}({\bf r},n) & = &
\frac{\lambda^{2}(v/\omega_{0})^{D+2s}}{L^{D}} \sum_{{\bf
    k}\neq0}\left|{\bf k}\right|^{2s}
\left[\frac{1-\cos(\omega_{k}\Delta)}{\omega_{k}^{2}}\right] \sin({\bf
  k}\cdot{\bf r})\sin(n\omega_{k}\Delta).
\label{eq:correlator-5}
\end{eqnarray}
\end{subequations}
\end{widetext}
Equation (\ref{eq:H}) can be interpreted as a statistical mechanics
Hamiltonian of a three-dimensional lattice of Ising variables. The
three dimensions are due to the the two-dimensional spatial lattice of
the qubits and the discrete ``time'' direction. The different
correlation functions originated from the bosonic model produce
interactions between these Ising variables that can be long or short
ranged.

Using this notation, the logical qubit fidelity can be cast as
\begin{equation}
  \mathcal{F} = \frac{\sum^{\prime}_{\bar{\sigma},\bar{\tau}}
    e^{-{\cal H}} \prod_l \left\langle \bar{\tau}_{l} \right| {\cal
      J}_{l}\, G \left| \bar{\tau}_{l} \right\rangle \left\langle
    \bar{\sigma}_{l}\right| {\cal J}_{l}\, G\left| \bar{\sigma}_{l}
    \right\rangle} {\sum_{\bar{\sigma},\bar{\tau}} e^{-{\cal H}}
    \prod_l \left\langle \bar{\tau}_{l} \right| {\cal J}_{l}\, G\left|
    \bar{\tau}_{l} \right\rangle \left\langle \bar{\sigma}_{l} \right|
        {\cal J}_{l}\, G \left| \bar{\sigma}_{l} \right\rangle}.
  \label{eq:Fidelity}
\end{equation}
There are two aspects to consider when analyzing this
expression. First, the sum over the Ising variables in the numerator
is constrained to the positive stars due to the projector $G$. Hence,
not all Ising configurations of three dimensional lattice contribute
to the sums. Second, the ``energy cost'' imposed by ${\cal H}$ assign
different weights to the terms of the sum. Because of exchange
symmetry $\sigma \leftrightarrow \tau$, these contributions are always
real (but very difficult to evaluate).

In Sec. \ref{sec:Super-ohmic-environment-with} we evaluate the fidelity
 for a particularly set
of environment parameters that allow for an analytical solution.

\section{Superohmic environment with $s=1/2$}
\label{sec:Super-ohmic-environment-with}

A very interesting case to consider has $s=1/2$ and $D=2$. This
corresponds to a superohmic environment where analytical expressions
for the correlations functions in Eqs. (\ref{eq:correlators}) can be
easily written by imposing the continuum limit
\cite{PhysRevA.88.012336}
\[
\frac{\left(2\pi\right)^{2}}{L^{2}} \sum_{\mathbf{k}} \rightarrow
\int_{0}^{\Lambda} \rho\, d\rho \int_{0}^{2\pi}d\theta.
\]
As usual when dealing with superohmic environments, some values for
the variables ${\bf r}$ and $n$ in Eqs. (\ref{eq:correlators}) can
lead to ultraviolet diverging contributions. The leading divergent
terms are linearly proportional to the ultraviolet cutoff,
\begin{equation}
F_{1}(0,0) \approx \frac{\lambda^{2}v\Lambda}{2\pi\omega_{0}^{3}},
\end{equation}
and
\begin{equation}
F_{1}\left(0,1\right) \approx -
\frac{\lambda^{2}v}{2\pi\omega_{0}^{3}} \frac{\Lambda}{2},
\end{equation}
with the remaining terms giving subleading contributions diverging
with the cutoff or no divergence at all. Hence, for a large
environmental cutoff, in leading order, it is a good approximation to
simplify Eq. (\ref{eq:H}) to
\begin{eqnarray}
\mathcal{H} & \approx & \frac{J}{4} \sum_{n=0}^{N-1} \sum_{\mathbf{r}}
\Big[ \left( \tau_{\mathbf{r},n} -\sigma_{\mathbf{r},n} \right) \left(
  \tau_{\mathbf{r},n} -\sigma_{\mathbf{r},n}\right) \\ & &
  +\ \frac{1}{2} \left( \tau_{\mathbf{r},n} - \sigma_{\mathbf{r},n}
  \right) \left( \tau_{\mathbf{r},n+1} - \sigma_{\mathbf{r},n+1}
  \right) \Big],
\end{eqnarray}
where $J = \frac{2\lambda^{2}v}{\pi\omega_{0}^{3}} \Lambda$.

We can now introduce the spin-1 operator
\begin{equation}
S_{{\bf r},n} \equiv \frac{1}{2}(\tau_{{\bf r},n}-\sigma_{{\bf r},n}),
\end{equation}
and write
\begin{equation}
\mathcal{H} \approx J \sum_{n=0}^{N-1} \sum_{\mathbf{r}} \left(
S_{{\bf r},n}^{2} + \frac{1}{2} S_{{\bf r},n}\, S_{{\bf r},n+1} \right),
\label{eq:spin-1-chain}
\end{equation}
which can be interpret as the Hamiltonian of decoupled spin-1 chains
along the ``time'' direction of the statistical mechanics model
associated to the fidelity calculation.

The first term in Eq. (\ref{eq:spin-1-chain}), $S_{{\bf r},n}^{2}$, is
a ``zero-field splitting'' (also known as on-site anisotropy) and
competes with the exchange coupling,
$S_{{\bf r},n}S_{{\bf r},n+1}$. The physics of this statistical
mechanical model is well known \cite{Capel1966}, and for the numerical
prefactors on the right-hand side of Eq. (\ref{eq:spin-1-chain}) there
is no phase transitions and no magnetic ordering. This means that the
qubit configurations between QEC cycles do not energetically constrain
each other in the calculation of the evaluating the fidelity through
Eq. (\ref{eq:Fidelity}). Hence, we can find the threshold by
discussing the critical coupling in a single QEC cycle. This result
provides a more rigorous justification for the usual assumption that
for a superohmic environment memory and correlations induced by the
environment can be neglected in the evaluation of the threshold.

The fidelity can be evaluated as
\begin{equation}
  \mathcal{F} = \frac{\sum^{\prime}_{\bar{\sigma},\bar{\tau}}
    \prod_{l} e^{-{\cal H}_{l}} \left\langle \bar{\tau}_{l} \right|
         {\cal J}_{l}\, G \left| \bar{\tau}_{l} \right\rangle
         \left\langle \bar{\sigma}_{l} \right| {\cal J}_{l}\, G \left|
         \bar{\sigma}_{l} \right\rangle}
          {\sum_{\bar{\sigma},\bar{\tau}} \prod_{l}e^{-{\cal H}_{l}}
            \left\langle \bar{\tau}_{l} \right| {\cal J}_{l}\, G
            \left| \bar{\tau}_{l} \right\rangle \left\langle
            \bar{\sigma}_{l} \right| {\cal J}_{l}\, G \left|
            \bar{\sigma}_{l} \right\rangle},
  \label{eq:Fidelity-1}
\end{equation}
where the single-chair Hamiltonian is given by
\begin{equation}
{\cal H}_{l} = J\, S_{{\bf r},l}^{2}.
\label{eq:H-slice}
\end{equation}

\begin{figure}
\begin{centering}
\includegraphics[scale=0.36]{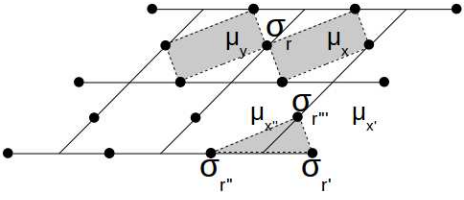}
\par\end{centering}
\caption{Illustration of mass field variables in the bulk and in the
  boundary for the spin variable $\sigma$.}
\label{fig:mass-fields}
\end{figure}

Although there is no explicit interaction between the spin-1
variables, the constrain of positive stars make the evaluation of the
fidelity nontrivial. The simplest method to deal with the constraint is
to use ``mass field variables''
\cite{PhysRevA.90.042315,arXiv:1606.0316}. The mass field $\mu$ is an
Ising variable defined in the center of each plaquette. The variable
$\sigma_{{\bf r}}$ in the link between two plaquettes is written as
the product of the mass field of each plaquette. For example, for bulk
sites in Fig. \ref{fig:mass-fields},
\begin{equation}
\sigma_{\mathbf{r},l} = \mu_{\mathbf{x},l}\, \mu_{\mathbf{y},l}
\label{eq:mass-fields-bulk}
\end{equation}
and
\begin{equation}
\tau_{\mathbf{r},l} = \nu_{\mathbf{x},l}\, \nu_{\mathbf{y},l}.
\end{equation}
We can now replace the sum over the constrained $\sigma$ and
$\tau$ variables for an unconstrained sum over $\mu$ and $\nu$
variables.

The only issue remaining are the top and bottom lattice boundaries
lattice. Stars located at the boundary are formed by only three
qubits, but because of the positive stars constraint the mass field
variables at these boundaries always assume the same value
\cite{PhysRevA.90.042315},
\begin{equation}
\mu_{\mathbf{p}',l}\, \sigma_{\mathbf{r}',l} = \mu_{\mathbf{p}'',l}\,
\sigma_{\mathbf{r}'',l} : = \alpha_{l} = \pm 1
\label{eq:mass-fields-bound}
\end{equation}
and
\begin{equation}
\nu_{\mathbf{p}',l}\, \tau_{\mathbf{r}',l} = \nu_{\mathbf{p}'',l}\,
\tau_{\mathbf{r}'',l} : = \beta_{l} = \pm 1.
\end{equation}
Using the new variables, Eq. (\ref{eq:H-slice}) can be rewritten as
\begin{eqnarray}
\mathcal{H}_{l} & = & -\frac{J}{4} \Bigg(
\sum_{\left\langle\mathbf{p},\mathbf{p'}\right\rangle\in\text{bulk}}
\nu_{\mathbf{p},l}\, \nu_{\mathbf{p'},l}\, \mu_{\mathbf{p},l}\,
\mu_{\mathbf{p'},l} \nonumber \\ & &
+\ \sum_{\mathbf{p}\in\text{boundary}} \alpha_{\mathbf{p},l}\,
\nu_{\mathbf{p},l}\, \mu_{\mathbf{p},l} \Bigg),
\label{eq:mass-field-model}
\end{eqnarray}
with no restrictions on the values of the mass field variables. We can
further simplify the problem by noticing that the change of variables
\begin{equation}
s_{\mathbf{p},l} = \nu_{\mathbf{p},l}\, \mu_{\mathbf{p},l}
\end{equation}
maps Eq. (\ref{eq:mass-field-model}) into a square lattice Ising model
with boundary fields \cite{arXiv:1606.0316}
\begin{equation}
\mathcal{H}_{l} = -\frac{J}{4} \left(\sum_{\left\langle
  \mathbf{p},\mathbf{p'}\right\rangle \in \text{bulk}}
s_{\mathbf{p},l}\, s_{\mathbf{p'},l} + \alpha_{l}\, \beta_{l}
\sum_{\mathbf{p}\in\text{boundary}} s_{\mathbf{p},l} \right).
\end{equation}

Following our previous work on this model and on the surface code
fidelity calculation \cite{PhysRevA.90.042315,arXiv:1606.0316} and
using the Onsager solution \cite{PhysRev.65.117}, we know that this
model has a second-order phase transition at the critical coupling
\begin{equation}
J_{c} = 2 \ln \left( 1 + \sqrt{2} \right).
\end{equation}
Thus, for $J>J_{c}$ the Ising model is in its ordered phase and the
fidelity is smaller than unity. Conversely, for $J<J_{c}$ the Ising
model is in its paramagnetic phase and the fidelity is unity
\cite{PhysRevA.90.042315,arXiv:1606.0316}. Thus, the critical coupling
constant for the surface code in this superohmic environment is
\begin{equation}
\lambda_{c} \approx 2 \sqrt{\frac{\pi\omega_{0}^{3}}{v\Lambda} \ln
  \left( 1+\sqrt{2} \right)}.
\label{eq:threshold}
\end{equation}
%
 
\section{Summary and Conclusions}
\label{sec:Conclusions}

There is a large body of theoretical work where the efficiency of QEC
is analyzed. One of the most common methods employed is the use of the
operator norm for the interaction Hamiltonian. However, this method is
inapplicable to models with diverging norms, such as the spin-boson
\cite{Terhal:2005:012336,AGP06,AKP06}. An alternative method is to use
a master equation for the quantum evolution of the reduced density
matrix, $\rho_{r}$ \cite{Weiss:1999:1}. Even though the master
equation formalism is very general, some very stringent assumptions
must be made even in the simplest cases in order to obtain workable
expressions. In addition, the initial step of the formalism is to
integrate the environmental degrees of freedom. Hence, it explicitly
precludes any correlations induced by the environment between qubits
at different QEC steps. In this paper we followed a third route: the
Feynman-Vernon influence functional formalism \cite{Weiss:1999:1}.

We followed the full quantum evolution of the system and the
environment and only at the end of the QEC evolution we traced the
environment. For a single QEC step, both the influence functional
formalism and a well-performed Lindblad description are expected to
yield similar results. However, due to the syndrome extraction
procedure, this equivalence may not hold when many QEC cycles are
considered.

For a pure bit-flip model, it was possible to write the exact time
evolution operator and then obtain an exact closed form for a logical
qubit fidelity after an arbitrary number of QEC cycles,
Eq. (\ref{eq:Fidelity}). Even for this simple pure bit-flip model the
expression is daunting. For the surface code it corresponds to a
three-dimensional lattice of interacting Ising variables that are
constrained by the nature of the QEC code. This result can be
understood in terms of a fictitious statistical mechanical problem
\cite{NMB07,Novais:2008:012314}. In this language, the QEC threshold
corresponds to a ``phase transition'', where the system-environment
coupling is mapped onto an effective inverse temperature. For a very
small coupling (corresponding to a high-temperature phase) the
statistical mechanical problem is in its disorder phase and QEC is able
to keep the fidelity equal to unity. However, for large couplings
(corresponding to a low-temperature phase) the qubits can ``order''
and the fidelity becomes smaller than unity.

We specialized the calculation for the superohmic case with $s=1/2$ and
a two-dimensional environment. Superohmic environments are plagued with
ultraviolet divergencies in some of the correlation functions that
yield interactions between qubits. In the particular example of
$s=1/2$ the leading diverging terms have a linear dependence with the
ultraviolet cutoff. Although we peformed an explicit calculation for
this particular value of $s$, similar results also hold for any $s>0$
case. Therefore, the discussions of Sec.
\ref{sec:Super-ohmic-environment-with} can be extended for all
superohmic cases.

In Sec. \ref{sec:Super-ohmic-environment-with} we demonstrated that
the statistical mechanical problem defined by Eq. (\ref{eq:Fidelity})
can be simplified in the superohmic case to an array of spin-1 chains,
Eq. (\ref{eq:spin-1-chain}). The particular set of parameters that
emerged from the calculation tells us that the array is in a phase
where the ``zero-field splitting'' term is dominant. Hence, the model
can be further simplified and only this ``zero-field splitting'' term
need to be kept. This analysis justifies the use of stochastic error
models in the discussion of the QEC threshold for superohmic
environments. The model has an analytical solution, allowing us to
find an expression for the critical coupling where the
threshold takes place, Eq. (\ref{eq:threshold}).

In summary, we provided and exact expression for the fidelity of a
logical qubit in the surface code in the presence of a bosonic
environment after an arbitrary number of QEC steps. We demonstrated
that for superohmic environments the use of stochastic models is fully
justifiable, thus confirming in an explicitly example an old
conjecture of the literature \cite{NMB10}.

\section{Acknowledgments}

Daniel L\'opez acknowledges CNPq for financial support. A.O.C. and
E. Novais wish to thank CNPq and FAPESP through the initiative
INCT-IQ. This work was supported in part by the FAPESP Grant
2014/26356-9 and the NSF Grant CCF-1117241.



\end{document}